\documentclass[preprint,aps,showpacs,superscriptaddress]{revtex4}
\usepackage{graphicx}
\usepackage{dcolumn}
\usepackage{amsfonts, amssymb}

\newcommand{\m}{\mathbb}
\newcommand{\un}{\underline}
\newcommand{\be}{\begin{equation}}
\newcommand{\ee}{\end{equation}}

\begin{document}

\title{Isospin mixing effects in low-energy $\bar{K}N - \pi \Sigma$ interaction}

\author{J. R\'{e}vai}
\affiliation{Research Institute for Particle and
Nuclear Physics, H-1525 Budapest, P.O.B. 49, Hungary}

\author{N.V. Shevchenko\footnote{Corresponding author:
shevchenko@ujf.cas.cz}}
\affiliation{Nuclear Physics Institute, 25068 \v{R}e\v{z}, Czech Republic}

\date{\today}

\begin{abstract}
New strong coupled-channel $\bar{K}N - \pi \Sigma$ potential, reproducing all existing
experimental data and suitable for using in an accurate few-body calculations,
is constructed. Isospin breaking effects of direct inclusion of the Coulomb interaction
and using of physical masses in calculations are investigated. The $1 s$ level
shift and width of kaonic hydrogen, consistent with the scattering data, was
obtained and the corresponding exact strong $K^- p$ scattering length was calculated.
One- and two-pole form of $\Lambda(1405)$ resonance was considered.
\end{abstract}

\pacs{13.75.Jz, 11.80.Gw, 36.10.Gv}
\maketitle

\section{Introduction}

Kaonic atoms and, especially, possibility of formation of kaonic nuclear
clusters attracted large interest recently. For investigation
of these systems it is necessary to know the basic $\bar{K}N$ interaction,
which is strongly connected with $\pi \Sigma$ and other channels.

Different theoretical models were used for constructing antikaon-nucleon
interaction. All these models can be separated in two groups: ``stand-alone''
potentials only fitting two-body data and potentials to be used in future
(few- or many-body) calculations.

To the first group belong very popular in our days potentials based on
chiral Lagrangian. The method consists of constructing a potential which
gives amplitudes equivalent to those derived from an effective chiral
lagrangian. Such potentials have many channels, including energetically
closed near $\bar{K}N$ threshold ones.
The most recent example is a model constructed in~\cite{Borasoy_KN,Borasoy_aKp}.
It is good in reproducing the antikaon-nucleon experimental data, however, due to
its unhandiness the potential cannot be used in few- or many-body calculations.

On the other hand, effective potentials used in approximate few-body
calculations are too simple for proper describing of all properties of $\bar{K}N$ system.
In most cases a one-channel (effective) optical potential is used.
For example, $\bar{K}N$ potential, used in~\cite{ay} for calculating of deeply-bound
kaonic nuclear states, is an energy-independent optical potential. It was constructed
in such a way that it corresponds to the elastic part of a coupled-channel
phenomenological $\bar{K}N - \pi \Sigma - \pi \Lambda$
potential. However, already the coupled-channel potential is too simple.
One more example is a recent work~\cite{2bWeise}, where a potential for further use
in a few-body calculation was derived. It is once more an effective energy-dependent
optical potential by construction: it reproduces the elastic part of an effective chiral
coupled-channel model.

Two-body optical potential could be equivalent to the original coupled-channel
ones.  For separable potentials it is possible to construct exact optical potential,
but even an exact optical potential properly describes only the elastic part of the whole
system. Moreover, introducing such ``good'' effective optical potential into
$N>2$ equations does not guarantee proper
description of all inelastic effects taking place in a few- or many-body system.

The inelastic effects are especially important for the antikaon-nuclear systems,
because $\bar{K}N$ interaction is strongly coupled to the $\pi \Sigma$ channel
through $\Lambda(1405)$ resonance. However, the nature of the
resonance is a separate question. A usual assumption is that $\Lambda(1405)$ is a
resonance in $\pi \Sigma$ and a quasi-bound state in $\bar{K}N$ channel.
There is also an assumption suggested by a chiral model, that the bump, which is usually
understood as $\Lambda(1405)$ resonance, is an effect of two poles
(see e.g.~\cite{2L1405},~\cite{MagasOset}). 
Some challenge to the two pole model was put by the recent experiment
at COSY-J\"{u}lich~\cite{Zychor}, but a subsequent theoretical paper~\cite{geng}
seems to reproduce the experiment on the basis of the two pole model.

Other sources of experimental data about $\bar{K}N$ interaction are also non-precise,
old or controversial. The data on cross-sections of elastic and inelastic scattering
with $K^- p$ in the initial state are rather old with rather large errors, while
threshold branching ratios of $K^-p$ scattering were measured more accurately.

Another source of knowledge about $\bar{K}N$ is kaonic hydrogen atom. Several
experiments were performed for measuring $1s$ level
shift caused by strong $\bar{K}N$ interaction. The two recent ones are
KEK~\cite{KEK1s} and DEAR~\cite{DEAR1s} results. More
recent DEAR value of $1s$ level shift and width significantly differs from the older
KEK result, it has smaller errors, but is inconsistent with the scattering $K^- p$
data as was shown in~\cite{Borasoy_KN,Borasoy_aKp}.

Moreover, there is a problem common for both experimental papers: they present
a $K^- p$ scattering length following from the measurements as an ``experimental value''.
However, $a_{K^- p}$ values in~\cite{KEK1s} and~\cite{DEAR1s} were
obtained using Deser-Trueman (DT) formula~\cite{Deser}, while in many papers
(among them in~\cite{Revai} for several one-channel model potentials) it was shown,
that the approximate formula has poor accuracy,
in particular for the $\bar{K}N$ interaction. There are several papers, introducing
different corrections to DT, nowadays the most popular is a
formula from~\cite{Ruzecky}. Undoubtedly, the corrected formula~\cite{Ruzecky}
has the same advantage as original DT~\cite{Deser} one: it is a
model-independent relation between scattering length and atomic level shift and width.
Its accuracy can be checked within a potential model where exact calculations
are feasible.

Since the measured value is the $1s$ level shift and width
(and not the $K^- p$ scattering length) we decided to construct a
phenomenological coupled-channel potential, reproducing kaonic hydrogen's level
shift and width without intermediate reference to $a_{K^- p}$.
It is clear that for reproducing the level shift of kaonic hydrogen it is
necessary to include Coulomb interaction into equations directly, which beaks isospin
symmetry.
As far as we know, the only attempt to do the same was performed in~\cite{Cieply}.
The authors used their own method for calculating kaonic atomic state with separable
chiral-based strong part of the potential and tried to reproduce DEAR data.
However, the resulting potential~\cite{Cieply} provides too large width
$\Gamma$ of $1s$  kaonic hydrogen level in comparison with DEAR values,
moreover, there are problems with reproducing $\Lambda(1405)$ resonance.
The first version of our $\bar{K}N - \pi \Sigma$ potential reproducing $1s$
level shift instead of $K^- p$ scattering length with direct inclusion of the
Coulomb interaction, and the corresponding three-body
$\bar{K}NN - \pi \Sigma N$ calculation using the obtained potential
was presented in~\cite{Pisa_proc_my}.

There is one more approximation which is widely used in theoretical models, namely,
neglecting the mass difference in iso-multiplets. However, the difference of masses
between proton and neutron and $K^-$ and $\bar{K}^0$ is a physical fact. Besides,
the effect of taking the mass difference into account is especially important in
the near-threshold $\bar{K}N$ region. Using the physical masses in the calculations
is one more isospin symmetry breaking effect, taken into account in the paper.

Thus, our aim is to construct phenomenological coupled-channel
$\bar{K}N - \pi \Sigma$ potential, which within the limits of the possible
simultaneously reproduce all experimental data: the level shift
and width of kaonic hydrogen $1s$ level (KEK or DEAR values), $K^- p$ threshold
branching ratios, elastic and inelastic $K^- p$ scattering, and $\Lambda(1405)$
resonance in one- or two-pole form.
We directly include such isospin breaking effects as Coulomb interaction and using
the physical masses of particles in the calculations. The corresponding $T$-matrix
should be suitable for
using in an accurate few-body (for example, three-body coupled-channel Faddeev)
calculation.

\section{Formulation of the problem}

Our non-relativistic Hamiltonian has the form
\begin{equation}
H = H^0 + V^c + V^s
\end{equation}
with $H^0$ being the kinetic energy plus the threshold energy of particle pairs,
$V^c$ and $V^s$ denote their Coulomb and strong interaction, respectively. The
transition matrix for the problem defined by this Hamiltonian can be written as
\begin{equation}
\label{Tba}
T_{ba} = T_{ba}^{c} + T_{ba}^{sc},
\end{equation}
where $T_{ba}^{c}$ is the pure Coulomb transition matrix, while $T_{ba}^{sc}$ is
the so called Coulomb-modified strong transition matrix, defined as
\begin{equation}
\label{Tbasc}
T_{ba}^{sc} = \langle \Phi_b^{c(-)} | V^s | \Psi_{a}^{(+)} \rangle .
\end{equation}
Here $|\Phi_b^{c(\pm)}\rangle$ is a Coulomb scattering state labeled by the
final state index $b$, while  $|\Psi_a^{(+)}\rangle$ denotes the total scattering
state, corresponding to the initial state labeled $a$ and satisfying the
Lippmann-Schwinger equation
\begin{equation}
\label{psiLS}
|\Psi_{a}^{(+)}\rangle =
|\Phi_a^{c(+)} \rangle + G^c(E+i\varepsilon) V^s |\Psi_{a}^{(+)} \rangle
\end{equation}
with the Coulomb Green's function
\begin{equation}
G^c(z) = (z-H^0-V^c)^{-1}\ .
\end{equation}
For a separable strong potential taken as
$V^s=|g\rangle\lambda\langle g|$ the $T_{ba}^{sc}$ matrix~(\ref{Tbasc}) has a form
\begin{equation}
\label{Tfi}
T_{ba}^{sc} = \langle \Phi_b^{c(-)} |g \rangle
(\lambda^{-1} - \langle g|G^c(E+i\varepsilon) |g\rangle )^{-1}
\langle g|\Phi_a^{c(+)}\rangle\ .
\end{equation}
For sufficiently simple form-factors $|g\rangle$ the matrix elements of the Coulomb
Green's function $\langle g | G^c(E+i\varepsilon) | g\rangle$ together with the overlaps
$\langle g |\Phi_a^{c(\pm)} \rangle$ in Eq.(\ref{Tfi}) can be calculated analytically
(see e.g.~\cite{gGg_Coul1,gGg_Coul2,gGg_Coul3}).
The poles of the total $T_{ba}(z)$ matrix in this case are determined by the equation
\begin{equation}
\label{bound1cond}
\lambda^{-1} - \langle g| G^c(z) |g\rangle=0\ ,
\end{equation}
since it can be shown, that the poles of the pure Coulomb $T_{ba}^c$ matrix are canceled out
from Eq.(\ref{Tba}).

The non-relativistic description of transitions allowing for change of particle composition
is achieved by enlarging the Hilbert-space by adding to it a discrete ``particle composition''
index. In this case the operators and wave functions become matrices and vectors with respect
to this index. The details of the matrix formulation of Eqs.(\ref{Tbasc})--(\ref{bound1cond})
are described in the Appendix.

\section{Details of the calculation and the input}

In momentum representation the strong interaction matrix (A.16) can be written as:
\begin{equation}
\label{Vseprb}
V^s_{\m{I}_i,\m{I}_j}(k_{\m{\, I}_i},k_{\m{\, I}_j}) = \delta_{I(\m{I}_i),I(\m{I}_j)} \,
g_{\m{I}_i}(k_{\, \m{I}_i}) \lambda_{\m{I}_i,\m{I}_j} \, g_{\m{I}_j}(k_{\, \m{I}_j})
\end{equation}
with $g_{\m{I}_i}(k_{\, \m{I}_i}) = \langle\vec{k}_{\, \m{I}_i}|g_{\m{I}_i}\rangle$,
$\vec{k}_{\, \m{I}_i}$ being the relative momentum of the particles in $\m{I}_i$.
We use the $\hbar=c=1$ system of units, our plane waves are normalized as
$\langle\vec{k}|\vec{k}'\rangle=\delta(\vec{k}-\vec{k}')$. In this case the scattering
amplitude $f_{ba}$ is connected with $T_{ba}$ by:
\begin{equation}
f_{ba}=-(2\pi)^2\ \sqrt{\mu_a\mu_b}\ T_{ba} ,
\end{equation}
where $\mu_a$ ($\mu_b$) is the reduced mass of the particles in the initial (final) state.

We tried to reproduce simultaneously the following experimental data (A--D).

\subsection{$\Lambda(1405)$ resonance}

Mass $M_{\Lambda}$ and width $\Gamma_{\Lambda}$ of the $\Lambda(1405)$ resonance
according to the Particle Data Group~\cite{PDG} are:
\begin{equation}
\label{PDG_1405}
M_{\Lambda}^{\, \rm PDG} = 1406.5 \pm 4.0 \; {\rm MeV}, \quad
\Gamma_{\Lambda}^{\, \rm PDG} = 50.0 \pm 2.0 \; {\rm MeV}.
\end{equation}
Unlike to PDG, our $\Lambda(1405)$ is not a clear
$I=0$ state, but a mixture of $I=0$ and $I=1$ states. Having in mind
existing assumptions, we used two versions of  $\Lambda(1405)$'s ``nature'':
one- and two-pole ones.
For the one-pole form of $\Lambda(1405)$ we used Yamaguchi form-factors:
\begin{equation}
\label{1res_ff}
 g_{\m{\, I}_i}^{1pole}(k_{\m{\, I}_i}) = \frac{1}{(k_{\m{\, I}_i})^2 + (\beta_{\m{\, I}_i})^2}, \quad
 i = 1, \dots, 5.
\end{equation}
We assumed $\Lambda(1405)$ as a resonance in $\pi \Sigma$ and
a quasi-bound state in $\bar{K}N$ channel. So, calculation of~(\ref{boundcplcond}) was done
at physical sheet for $\bar{K}N$ and non-physical sheet for $\pi \Sigma$ channel.

For two-pole case we assumed that there are two resonances in $\pi \Sigma$ channel.
One of them, as before, originates from a bound state in $\bar{K}N$ channel,
the other one from a resonance in $\pi \Sigma$ channel (with $\bar{K}N - \pi \Sigma$
coupling switched off).
It is known that in a one-channel case a one-term separable potential with Yamaguchi
form-factors~(\ref{1res_ff}) and real strength parameters can not describe a resonance.
So, in order to have
a resonance in the uncoupled $\pi \Sigma$ channel, for two-pole $\Lambda(1405)$ case we used
$\pi \Sigma$ form-factors in the following form:
\begin{equation}
\label{2res_ffpi}
 g_{\m{\, I}_i}^{2pole}(k_{\m{\, I}_i}) = \frac{1}{(k_{\m{\, I}_i})^2 + (\beta_{\m{\, I}_i})^2} \,+\,
 \frac{s \, (\beta_{\m{\, I}_i})^2}{[(k_{\m{\, I}_i})^2 + (\beta_{\m{\, I}_i})^2]^2}, \quad
 i = 3, 4, 5.
\end{equation}
By this for the two-pole case we introduced one more parameter $s$. For the $\bar{K}N$ channel
here we used Yamaguchi form-factors:
\begin{equation}
\label{2res_ffK}
 g_{\m{\, I}_i}^{2pole}(k_{\m{\, I}_i}) = \frac{1}{(k_{\m{\, I}_i})^2 + (\beta_{\m{\, I}_i})^2}, \quad
 i = 1, 2.
\end{equation}
Both poles are once more situated at physical sheet for $\bar{K}N$ and non-physical sheet for
$\pi \Sigma$ channel.

\subsection{Kaonic hydrogen data}
The $K^- p$ atomic $1s$ level shift $\Delta E_{1s}$
and width $\Gamma_{1s}$ measured in the KEK experiment~\cite{KEK1s}
\begin{equation}
\Delta E^{KEK}_{1s} = -323 \pm 63 \pm 11 \; {\rm eV}, \quad
\Gamma^{KEK}_{1s} = 407 \pm 208 \pm 100 \; {\rm eV}
\end{equation}
and in the DEAR collaboration experiment~\cite{DEAR1s}
\begin{equation}
\Delta E^{DEAR}_{1s} = -197 \pm 37 \pm 6 \; {\rm eV}, \quad
\Gamma^{DEAR}_{1s} = 249 \pm 111 \pm 30 \; {\rm eV}
\end{equation}
differs from each other. We tried to reproduce both these
values within $1 \sigma$ interval.

We would like to stress, that in our approach there is no intermediate reference
to $K^- p$ scattering length when reproducing the level shift and the width.
Of course, after finding a set of potential
parameters we can calculate a strong scattering length, which
exactly corresponds to the obtained $1s$ level shift $\Delta E_{1s}$
and width $\Gamma_{1s}$. Due to the isospin symmetry breaking the formula
for the $a_{K^- p}$ differs from commonly used
$\frac{1}{2} \, (a_{\bar{K}N,I=0} + a_{\bar{K}N,I=1})$, since our $T$-matrix
has non-diagonal elements between $I=0$ and $I=1$ states.

We mention here, that energies of atomic (kaonic hydrogen $1s$ level)
and nuclear (one- and two-pole $\Lambda(1405)$) states are obtained from the
same system of equations~(\ref{boundcplcond}). The second remark concerns the
origin of the resonances. All our resonances are poles on the corresponding
sheet of the complete problem. Since our formula~(\ref{boundcplcond}) was
obtained by solving dynamical equations, the resonances
can be rightly called dynamically generated ones.

\subsection{Scattering data}
Elastic and inelastic total cross sections with $K^- p$
in the initial state were measured in~\cite{Kp2exp,Kp3exp,Kp4exp,Kp5exp,Kp6exp}
(we did not take into consideration data from~\cite{Kp1exp} with huge error bars).
It is interesting,
that there are no comments about non-existence of the total elastic
cross-sections (except~\cite{Borasoy_KN} and~\cite{Borasoy_aKp})
due to the singularity of the pure Coulomb transition matrix $T_{ba}^c$
in~(\ref{Tba}), while the ``total elastic'' cross-sections are plotted by
almost every author of $\bar{K}N$ interaction models. Having Coulomb
interaction directly included into the calculations we could not ignore the
problem. We defined ``total elastic'' $K^- p$ cross-section following the
experimental works~\cite{Kp1exp, Kp2exp}.
Namely, the total cross-sections were obtained by integrating differential
cross-sections in the region $-1 \leq \cos{\theta} \leq 0.966$ instead of
$-1 \leq \cos{\theta} \leq 1$.

\subsection{Threshold branching ratios}
Three threshold branching ratios of $K^- p$ scattering were measured
rather accurately~\cite{gammaKp1, gammaKp2}. One of them is
\begin{equation}
\gamma = \frac{\Gamma(K^- p \to \pi^+ \Sigma^-)}{\Gamma(K^- p \to
\pi^- \Sigma^+)} = 2.36 \pm 0.04 \, .
\end{equation}
We oriented on the medium value
\begin{equation}
\label{gamma_med}
\gamma = 2.36.
\end{equation}
The other two ratios $R_c$ and $R_n$, containing $K^- p \to \pi^0 \Lambda$
cross-sections,
\begin{eqnarray}
\label{Rc}
R_c &=& \frac{\Gamma(K^- p \to \pi^+ \Sigma^-, \pi^- \Sigma^+)}{\Gamma(K^- p \to
\mbox{all inelastic channels} )} = 0.664 \pm 0.011, \\
\label{Rn}
R_n &=& \frac{\Gamma(K^- p \to \pi^0 \Lambda)}{\Gamma(K^- p \to
\mbox{neutral states} )} = 0.189 \pm 0.015,
\end{eqnarray}
could not be used in a straightway because we did not include $\pi^0 \Lambda$
channel directly into our calculations. However, the effect of the channel
was effectively taken into account by allowing $\lambda^{1}_{\bar{K}N,\bar{K}N}$
parameter to have non-zero imaginary part (it significantly improved the
agreement with the experimental cross-sections). It is easy to find from the measured
$K^- p$ threshold branching ratios $\gamma$, $R_c$, and $R_n$, that relevant weight of
$\pi^0 \Lambda$ channel at $K^- p$ threshold among all possible
inelastic channels is approximately equal to $6 \%$.
So, the introduced imaginary part only slightly breaks unitarity in contrast
to what happens when a one-channel complex $\bar{K}N$ potential is used,
approximately accounting for the main inelastic $\pi \Sigma$ channel.

From existing $R_c$ and $R_n$ we constructed a new threshold branching ratio
\begin{equation}
 R_{\pi \Sigma} =
 \frac{\Gamma(K^- p \to \pi^+ \Sigma^-)+\Gamma(K^- p \to \pi^- \Sigma^+)}{
 \Gamma(K^- p \to \pi^+ \Sigma^-) + \Gamma(K^- p \to \pi^- \Sigma^+) +
                                   \Gamma(K^- p \to \pi^0 \Sigma^0) } \,.
\end{equation}
From definitions of $R_c$ and $R_n$~(\ref{Rc}),~(\ref{Rn})  using experimental data
we obtained for the $R_{\pi \Sigma}$ an ``experimental'' value
\begin{equation}
 R_{\pi \Sigma} =  \frac{R_c}{1-R_n \, (1 - R_c)} \, = \, 0.709 \pm 0.011 \,.
\end{equation}
We tried to reproduce the medium value
\begin{equation}
\label{RpiSig_med}
 R_{\pi \Sigma} =  0.709.
\end{equation}

The formulae~(\ref{Tfi_matrix}) and~(\ref{boundcplcond}) allow us to find
parameters $\lambda_{\m{I}_i,\m{I}_j}$, $\beta_{\m{I}_i}$ (and $s$)
of our potentials in both one-pole and two-pole cases, which reproduce
these experimental quantities. All our parameters,
except $\lambda^{1}_{\bar{K}N,\bar{K}N}$, are real.

\section{Results and discussion}

We started the calculations with inclusion of the Coulomb interaction and
using physical masses in both $\bar{K}N$ and $\pi \Sigma$ channels.
However, it turned out, that the effects are small for the
$\pi \Sigma$ subsystem compared to those for the antikaon-nucleon
channel. It is understandable, since we are interested
in the energy region near $\bar{K}N$ threshold, where the mass difference
between $K^-$ and $\bar{K}^0$, $p$ and $n$ should manifest itself, at least, by the
existence of two close thresholds for $K^- p$ and $\bar{K}^0 n$
in contrast to the one threshold for $\bar{K}N$. Due to this we kept the Coulomb
potential in $K^- p$ subsystem and physical masses in
$\bar{K}N_{I=0}, \bar{K}N_{I=1}$ ($K^- p, \bar{K}^0 n$) channels, while in
$\pi \Sigma$ channels we used isospin averaged masses without the
Coulomb interaction.

In the case of averaged masses without Coulomb in $\pi \Sigma$
the $\pi \Sigma_{I=2}$ ($\m{I}_5$) channel is dynamically decoupled from the
other four channels. So, we can work in particle space of four dimensions,
corresponding to $\bar{K}N_{I=0}$, $\bar{K}N_{I=1}$ (or $K^- p, \bar{K}^0 n$),
$\pi \Sigma_{I=0}$, and $\pi \Sigma_{I=1}$ channels.
\begin{figure}[pb]
\centering
\includegraphics[width=0.95\textwidth]{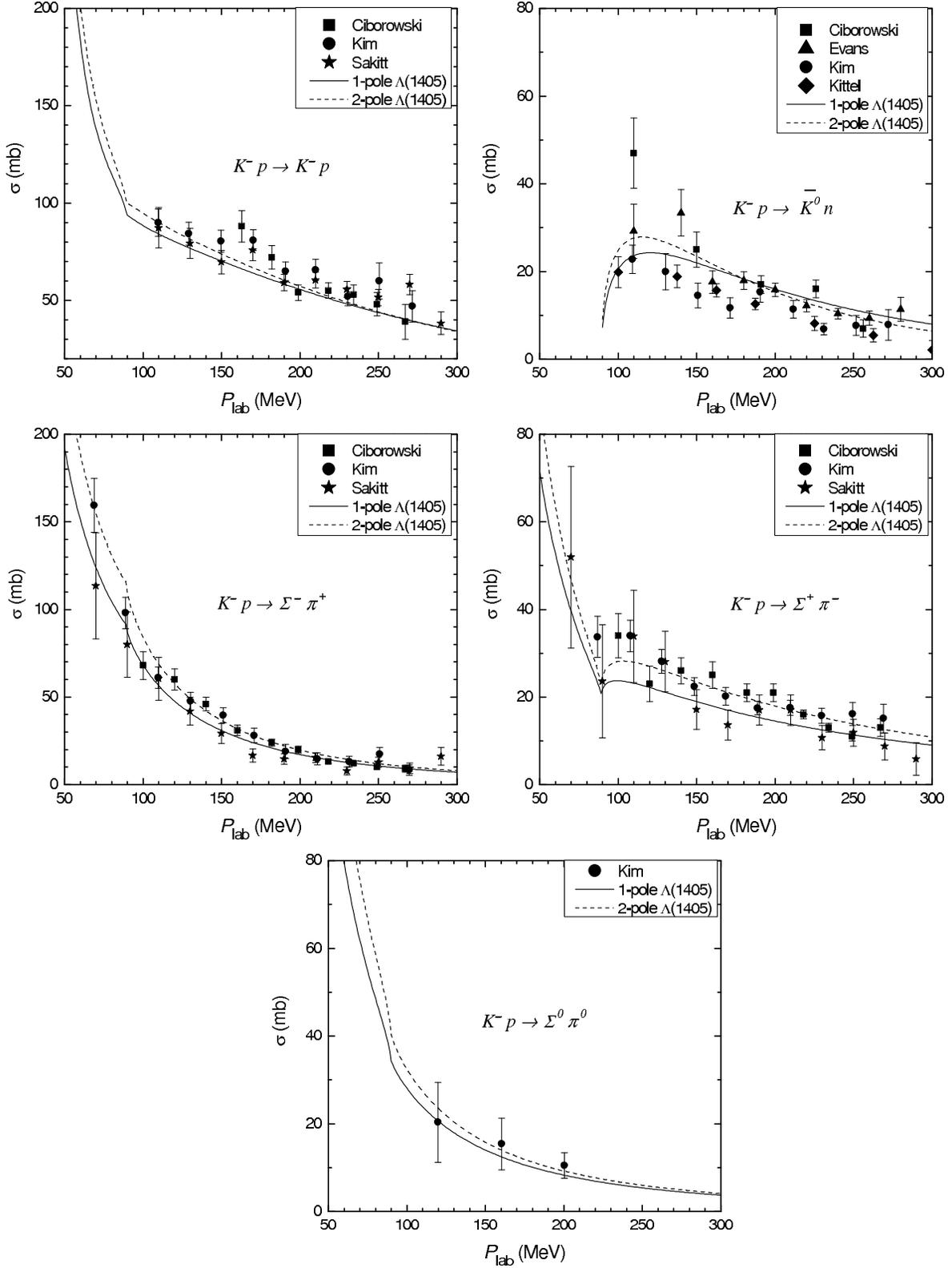}
\caption{Comparison of the obtained theoretical cross-sections
(lines) with experimental
data~\protect\cite{Kp2exp,Kp3exp,Kp4exp,Kp5exp,Kp6exp} (points).
Solid lines: calculation with 1-pole $\Lambda(1405)$ resonance,
dashed lines: calculation with 2-pole $\Lambda(1405)$ resonance.
\label{Five_plots}}
\end{figure}

We succeeded in obtaining parameters of the potentials with one- and
two-pole $\Lambda(1405)$ structure. The best set of the obtained parameters for
the one-pole $\Lambda(1405)$ is:
\begin{equation}
\begin{tabular}{c}
$\beta^{1 pole}_{\bar{K}N} = 3.4 {\rm \; fm}^{-1}$ \\
$\beta^{1 pole}_{\pi \Sigma} = 1.9 {\rm \; fm}^{-1}$
\end{tabular} \qquad
\underline{\underline{\Lambda}}^{1 pole}(\m{I}) = \left(
    \begin{tabular}{cccc}
    $-1.31$ & $0$ & $0.62$ & $0$ \\
    $0$ & $1.76 - i 0.24$ & $0$ & $1.90$ \\
    $0.62$ & $0$ & $0.18$ & $0$ \\
    $0$ & $1.90$ & $0$ & $1.24$
    \end{tabular}
\right)
\end{equation}
for the two-pole $\Lambda(1405)$ it is:
\begin{equation}
\begin{tabular}{c}
$\beta^{2 pole}_{\bar{K}N} = 3.2 {\rm \; fm}^{-1}$ \\
$\beta^{2 pole}_{\pi \Sigma} = 1.0 {\rm \; fm}^{-1}$ \\
$s = -0.87$
\end{tabular} \qquad
\underline{\underline{\Lambda}}^{2 pole}(\m{I}) = \left(
    \begin{tabular}{cccc}
    $-1.06$ & $0$ & $0.40$ & $0$ \\
    $0$ & $0.97 - i 0.11$ & $0$ & $1.13$ \\
    $0.40$ & $0$ & $-0.01$ & $0$ \\
    $0$ & $1.13$ & $0$ & $0.61$
    \end{tabular}
\right) \, .
\end{equation}
Here we assumed isospin-independence of the range parameters:
\begin{eqnarray}
\beta_{\m{\, I}_1} = \beta_{\m{\, I}_2} \equiv \beta_{\bar{K}N}, \\
\beta_{\m{\, I}_3} = \beta_{\m{\, I}_4} \equiv \beta_{\pi \Sigma}.
\end{eqnarray}

Our results for the cross-sections with best set of the obtained parameters
with one-pole and two-pole $\Lambda(1405)$ are presented in
Fig.~\ref{Five_plots}: the elastic $K^- p \to K^- p$ cross-section and
inelastic $K^- p \to \bar{K}^0 n$, $K^- p \to \pi^+ \Sigma^-$,
$K^- p \to \pi^- \Sigma^+$, and $K^- p \to \pi^0 \Sigma^0$ cross-sections
are compared with existing
experimental data~\cite{Kp2exp,Kp3exp,Kp4exp,Kp5exp,Kp6exp}. It is seen,
that both versions of the potential are equally good
in describing the experimental data within the experimental
errors. Due to this fact, unfortunately, it is not possible to give
preference  to one of the $\Lambda(1405)$ versions.

Other physical characteristics of the obtained
1-pole and 2-pole potentials are shown in Table~\ref{phys_char.tab}:
pole positions $z_1$ and $z_2$ (obviously, $z_2$ exists in 2-pole variant of the
potential only), $1s$ kaonic hydrogen level shift $\Delta E_{1s}$ and width
$\Gamma_{1s}$. Threshold branching ratios $\gamma$~(\ref{gamma_med}) and
$R_{\pi \Sigma}$~(\ref{RpiSig_med}) are reproduced exactly in both cases.
Having complete set of potential parameters it is possible to calculate
the strong $K^- p$ scattering length corresponding to the given $\Delta E_{1s}$
and $\Gamma_{1s}$ exactly. The $a_{K^- p}$ for both potentials are also shown
in the Table~\ref{phys_char.tab}.
\begin{center}
\begin{table}[ht]
\caption{Physical characteristics of the obtained
1-pole and 2-pole potentials: pole positions $z_1$ and $z_2$,
level shift $\Delta E_{1s}$ and width $\Gamma_{1s}$ of kaonic hydrogen,
and corresponding exact strong scattering length $a_{K^-p}$.
Threshold branching ratios~(\ref{gamma_med}) and~(\ref{RpiSig_med}) are
reproduced exactly.}
\label{phys_char.tab}
\begin{tabular}{ccc}
\hline \hline \noalign{\smallskip}
 & 1-pole $\Lambda(1405)$ & 2-pole $\Lambda(1405)$ \\
\noalign{\smallskip} \hline \noalign{\smallskip}
$z_1$ (MeV) & \qquad $1409 - i 32$  \qquad & \qquad
       $1412 - i 32$ \qquad\\
$z_2$ (MeV) & \qquad $-$  \qquad & \qquad
       $1380 - i 105$ \qquad\\
$\Delta E_{1s}$ (eV)   & $-396$ &  $-407$ \\
$\Gamma_{1s}$ (eV)     & $370$ &  $476$  \\
$a_{K^- p}$ (fm)  & $-1.07 + i 0.59$ & $-1.08 + i 0.76$ \\
\noalign{\smallskip} \hline \hline
\end{tabular}
\end{table}
\end{center}

\begin{figure}[hb]
\centering
\includegraphics[width=0.50\textwidth, angle=-90]{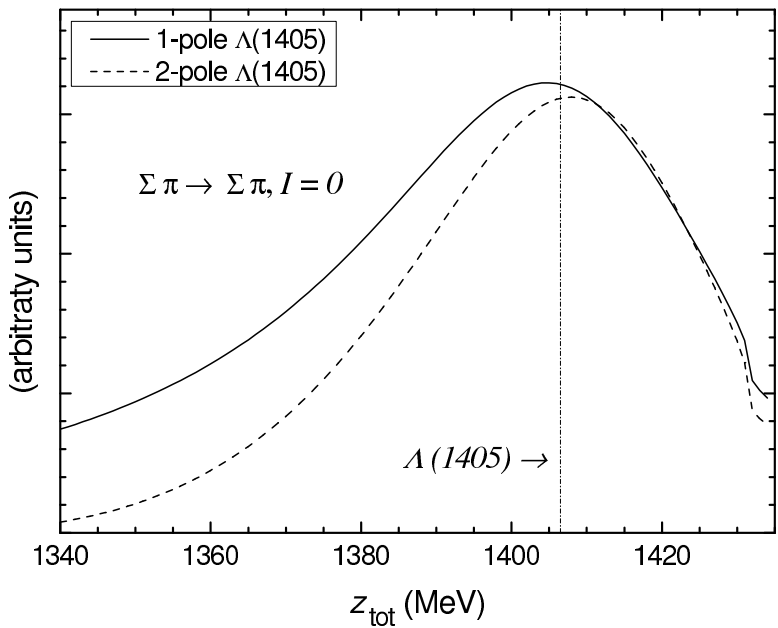}
\caption{Manifestation of $\Lambda(1405)$ resonance in $I=0$
$\pi \Sigma \to \pi \Sigma$ cross-section, solid line: calculation with
1-pole $\Lambda(1405)$, dashed line: calculation with 2-pole
$\Lambda(1405)$ resonance. The vertical line marks the medium PDG mass
$M^{PDG}_{\Lambda} = 1406.5$ MeV of the resonance~\cite{PDG}.
\label{Lambda1405}}
\end{figure}
The first pole positions $z_1$ for both versions of the potential have
close real parts and the same imaginary ones, however, all three numbers differ
from the PDG data for mass and width of $\Lambda(1405)$ resonance~(\ref{PDG_1405}).
The characteristics of the two poles $z_1$ and $z_2$ in the 2-pole $\Lambda(1405)$
version are the same as in~\cite{MagasOset}: one of them has less mass and larger
width, while the other is heavier with narrower width. However, the
positions of $z_1$ and $z_2$ differ from those in~\cite{MagasOset}.

\begin{figure}[hb]
\centering
\includegraphics[width=0.50\textwidth, angle=-90]{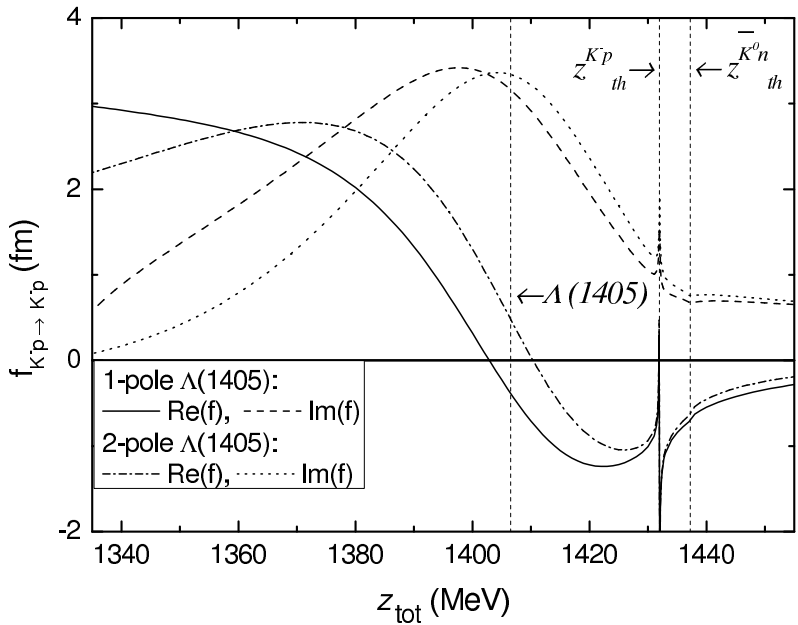}
\caption{Manifestation of $\Lambda(1405)$ resonance in
$K^- p \to K^- p$ amplitude below the threshold for both versions of the
potential. One-pole $\Lambda(1405)$: real (solid line) and imaginary
part (dashed line) of $f_{K^- p \to K^- p}$. Two-pole $\Lambda(1405)$:
real (dash-dotted line) and imaginary part (dotted line) of the amplitude.
The vertical lines marks the medium PDG mass
$M^{PDG}_{\Lambda} = 1406.5$ MeV of the resonance~\cite{PDG}, $K^- p$
and $\bar{K}^0 n$ thresholds.
\label{amplKp}}
\end{figure}
It is not absolutely clear, how to relate the obtained potentials to the shape of
the $\Lambda(1405)$ resonance. The experimental shape of the resonance is deduced
from missing mass experiments since direct $\pi \Sigma$ data are not available.
However, their relation to the pole structure of the two-body $T$-matrix
is not trivial and needs further investigation.
An example of the interpretation ambiguity is shown in
Fig.~\ref{Lambda1405}, where we demonstrate the manifestation of $\Lambda(1405)$
in a calculated isospin-zero elastic $\pi \Sigma$ cross-section.
It can be seen, that the maxima of
the resonances for the two versions of the potential lay on opposite sides
of the medium PDG value $M^{PDG}_{\Lambda} = 1406.5$ MeV~\cite{PDG}, while
both Re$(z_1)$ are larger, than $1406.5$ MeV (see Table~\ref{phys_char.tab}).

Another example is given in Fig.~\ref{amplKp}, where real and imaginary parts of
the elastic $K^- p$ amplitude for the two versions of the potential are depicted.
At the the resonance positions real parts of $f_{K^- p \to K^- p}$ have zeros
(situated at different, in respect to the medium PDG value, sides), while
imaginary parts have their maxima (at slightly lower energies). The Coulomb
singularities are seen almost at the $K^- p$ threshold.

\begin{figure}[ht]
\centering
\includegraphics[width=0.50\textwidth, angle=-90]{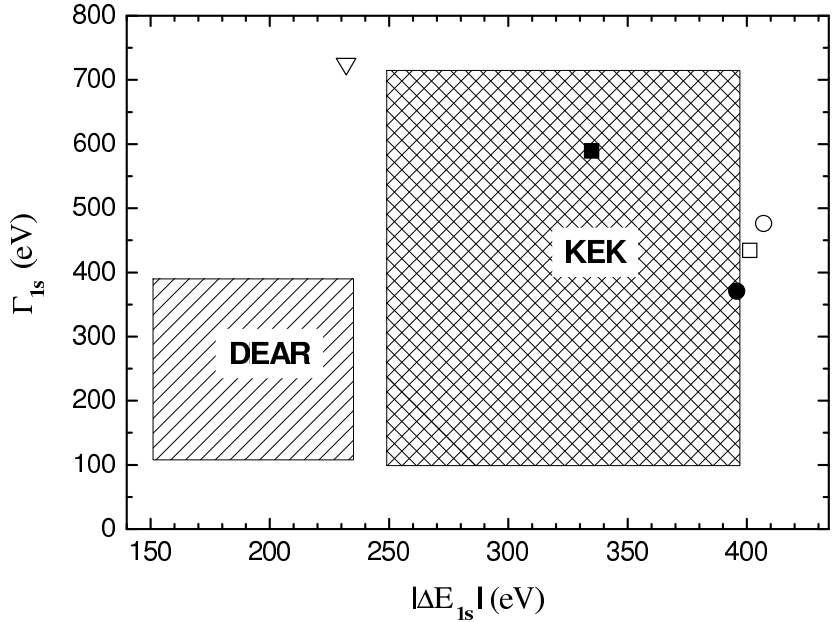}
\caption{DEAR and KEK $1 \sigma$ confidence region of kaonic hydrogen $1s$
level shift $|\Delta E|$ (absolute value) and width $\Gamma$. The
obtained theoretical results for the one-pole (solid circle) and two-pole
(empty circle) variants of the potential are shown. The results of other
theoretical models are also depicted: \cite{Borasoy_KN}~(solid square),
\cite{Borasoy_aKp}~(empty sqare), and \cite{Cieply}~(empty triangle).
\label{KEK_DEAR}}
\end{figure}

We plotted also the obtained parameters of kaonic hydrogen
($\Gamma_{1s},|\Delta E_{1s}|$), shown in Table~\ref{phys_char.tab}, together with
the experimental $1 \sigma$ regions of KEK and DEAR results, see Fig.~\ref{KEK_DEAR}.
It is seen, that obtained $\Delta E$ for the 1-pole version is situated inside
the KEK region, while for the 2-pole variant it is slightly outside. Both
values are close to each other, they definitely prefer the largest values of KEK
$|\Delta E|$. All our attempts to move the shift values to the DEAR region
led to drastic worsening of the agreement with the experimental cross-sections.
From this fact we do the same conclusion as did authors of~\cite{Borasoy_aKp}:
the DEAR data on kaonic hydrogen measurements are inconsistent with
the existing scattering data.

As for the widths, both are situated inside KEK $1\sigma$ limits,
while the 1-pole potential gives $\Gamma_{1s}$ also inside DEAR, closely to
its highest possible value. The important fact is that the obtained theoretical
values of $\Gamma_{1s}$ for the two versions of potentials have rather large
difference. But, unfortunately, the accuracy of KEK results does not allow to
make a unique selection between them.

For comparison we plotted also the results of other
theoretical models: \cite{Borasoy_KN}, \cite{Borasoy_aKp}, and \cite{Cieply}.
The first two ($\Gamma_{1s},|\Delta E_{1s}|$) values were obtained
from the $K^- p$ scattering lengths using corrected DT
formula~\cite{Ruzecky}, while the last one was calculated directly.
The chiral potential~\cite{Borasoy_aKp},
aiming to reproduce mainly the $K^- p$ scattering data, have the result
(corresponding to the best $a_{K^- p}$ value in the full approach)
impressively close to our, though the correctness of it is limited by
the corrected DT formula accuracy. The previous potential
of the same authors (version {\it ``u''}) have different ($\Gamma_{1s},|\Delta E_{1s}|$)
value, however, is also situated inside $1 \sigma$ KEK region. The result
of~\cite{Cieply} is far from all other theoretical values and outside both
experimental regions. The reason could be their
attempt to fit DEAR values simultaneously with the scattering data, which
turned out to be unsuccessful. It is an additional demonstration of
inconsistency of the DEAR results with the existing scattering data.

We see that both versions of our potential reproduce experimental
cross-sections equally well, by construction they exactly reproduce
threshold branching ratios $\gamma$ and $R_{\pi \Sigma}$. The obtained
values of the kaonic hydrogen level shift $\Delta E_{1s}$ in both versions
of $\Lambda(1405)$ resonance are close to each other. However, there is rather
large, more than $100$ eV, difference between the $K^- p$ widths
$\Gamma_{1s}$. Having in mind a forthcoming experiment SIDDHARTA~\cite{SIDDHARTA},
we hope, that the new experimental value will be close to one of our numbers
allowing to make a conclusion about the structure of $\Lambda(1405)$ resonance.

\begin{center}
\begin{table}[htb]
\caption{Isospin conserving $a^{cons}$ (fm) and non-conserving $a^{nonc}$ (fm)
constituents of the total $a$ (fm) $K^- p$ scattering length for one-pole
and two-pole versions of the potential.}
\label{cons_nonaKp.tab}
\begin{tabular}{lrr}
\hline \hline \noalign{\smallskip}
 & 1-pole $\Lambda(1405)$ & \quad 2-pole $\Lambda(1405)$ \\
\noalign{\smallskip} \hline \noalign{\smallskip}
$a^{cons}_{K^- p}$  &  $-1.0561 + i 0.6977$ &  $-0.9949 + i 0.8648$  \\
$a^{nonc}_{K^- p}$  &  $0.0139 + i 0.1077$  &  $0.0851 + i 0.1048$   \\
\hline
$a_{K^- p}$   &  $-1.07 + i 0.59$     &  $-1.08 + i 0.76$  \\
\noalign{\smallskip} \hline \hline
\end{tabular}
\end{table}
\end{center}
The combined effect of the exact inclusion of the Coulomb interaction and using
physical masses of the particles can be illustrated
by showing the isospin conserving and non-conserving parts of the $K^- p$
scattering length, see Table~\ref{cons_nonaKp.tab}. The constituents are defined as
\begin{eqnarray}
&{}& a_{K^- p}^{cons} \equiv \frac{1}{2} \, \left(a^{00}_{\bar{K}N} + a^{11}_{\bar{K}N} \right), \\
&{}& a_{K^- p}^{nonc} \equiv a^{01}_{\bar{K}N} = a^{10}_{\bar{K}N},
\end{eqnarray}
where $a^{II'}_{\bar{K}N}$ denotes the elastic strong (Coulomb is switched off)
$\bar{K}N$ on-shell amplitude
with initial (final) pair isospin I' (I) at the $K^- p$ threshold. The total
scattering length, also shown in Table~\ref{cons_nonaKp.tab}, is
\begin{equation}
a_{K^- p} = a_{K^- p}^{cons} - a_{K^- p}^{nonc} \,.
\end{equation}
It is seen, that real parts of non-conserving scattering lengths change the final
results only slightly, especially in the one-pole case, where $a_{K^- p}^{nonc}$
is $1 \%$  of  $a_{K^- p}$ (it is $8 \%$ for the two-pole variant). In contrast,
the imaginary parts change isospin conserving scattering lengths essentially,
the share of isospin non-conserving part is $18 \%$ for the one-pole and
$14 \%$ for the two-pole case. Thus, isospin breaking effects, taken into
account in our calculations, are important, especially for the strong $K^- p$
scattering length.

\begin{center}
\begin{table}[htb]
\caption{Kaonic hydrogen $1s$ level shift $\Delta E$ (eV) and width $\Gamma$ (eV),
corresponding to the obtained
scattering length: exact (this work), derived from Deser-Trueman formula~\cite{Deser},
and from corrected Deser formula~\cite{Ruzecky}.}
\label{exact_deser_ruz.tab}
\begin{tabular}{lrr}
\hline \hline \noalign{\smallskip}
 & 1-pole $\Lambda(1405)$ & \quad 2-pole $\Lambda(1405)$ \\
\noalign{\smallskip} \hline \noalign{\smallskip}
$\Delta E^{exact}_{1s}$   & $-396$ &  $-407$ \\
$\Gamma^{exact}_{1s}$     & $370$  &  $476$  \\
\hline
$\Delta E^{DT}_{1s}$ \cite{Deser}   & $-441$  &  $-445$  \\
$\Gamma^{DT}_{1s}$  \cite{Deser}    & $486$   &  $626$   \\
\hline
$\Delta E^{MRR}_{1s}$ \cite{Ruzecky}  & $-395$  &  $-411$  \\
$\Gamma^{MRR}_{1s}$ \cite{Ruzecky}    & $338$  &  $434$  \\
\noalign{\smallskip} \hline \hline
\end{tabular}
\end{table}
\end{center}
The differences of the exact $K^- p$ level shifts and widths, obtained
from our potentials, from results provided by approximate formulae are
demonstrated in Table~\ref{exact_deser_ruz.tab}.
The approximate DT~\cite{Deser} and corrected DT~\cite{Ruzecky} values
for the shift and width were obtained using our
exact scattering length given in the Table~\ref{phys_char.tab}.
It is seen, that DT formula~\cite{Deser}
gives very inaccurate result for both characteristics of kaonic atom:
the absolute value of the level shift and the width are overestimated. The
same result was obtained with several model one-channel complex $\bar{K}N$
potentials in~\cite{Revai}.
The widely used corrected Deser formula~\cite{Ruzecky} gives rather accurate
result for the shift, but underestimates the width of $1s$ level by $9-10 \%$.
\begin{center}
\begin{table}[ht]
\caption{One-pole potential: norms $\mathcal{N}$ of the
strong $z_1 = (1409.0- i 32.0)$~MeV and Coulomb
$z_c = (1431.9 - i 1.9 \times 10^{-4})$~MeV resonances.}
\label{norms1pole.tab}
\begin{tabular}{ccc}
\hline \hline \noalign{\smallskip}
& $z_1$  & $z_c$  \\
\noalign{\smallskip} \hline \noalign{\smallskip}
$\mathcal{N}_{(\bar{K}N)_{I=0}}$ & $1.288 - i 0.0792$  &  $0.500014 - i 0.000013$ \\
$\mathcal{N}_{(\bar{K}N)_{I=1}}$ & $0.0008 - i 0.0020$  &  $0.499986 + i 0.000012$  \\
$\mathcal{N}_{(\pi \Sigma)_{I=0}}$ & $-0.2885 + i 0.0810$  &  $\sim 10^{-7}$  \\
$\mathcal{N}_{(\pi \Sigma)_{I=1}}$  &  $-0.0001 + i 0.0002$  &  $\sim 10^{-7}$  \\
\noalign{\smallskip} \hline \noalign{\smallskip}
$\mathcal{N}_0$  &  $0.9993 + i 0.0018$  &  $0.500014 - i 0.000012$  \\
$\mathcal{N}_1$  &  $0.0007 - i 0.0018$  &  $0.499986 + i 0.000012$  \\
\noalign{\smallskip} \hline \noalign{\smallskip}
$\mathcal{N}_{K^- p}$  &  $0.681 - i 0.061$  &  $0.999994 + i 2.8 \times 10^{-6}$  \\
$\mathcal{N}_{\bar{K}^0 n}$  &  $0.608 - i 0.020$  &  $\sim 10^{-6}$  \\
$\mathcal{N}_{\pi^- \Sigma^+}$  &  $-0.101 + i 0.031$  &  $\sim 10^{-7}$  \\
$\mathcal{N}_{\pi^0 \Sigma^0}$  &  $-0.096 + i 0.027$  &  $\sim 10^{-7}$  \\
$\mathcal{N}_{\pi^+ \Sigma^-}$  &  $-0.092 + i 0.023$  &  $\sim 10^{-7}$ \\
\noalign{\smallskip} \hline \hline
\end{tabular}
\end{table}
\end{center}

\begin{center}
\begin{table}[ht]
\caption{
Two-pole potential: norms $\mathcal{N}$ of the
strong $z_1 = (1411.9 - i 32.0)$~MeV, $z_2 = (1380.0 - i 105.0)$~MeV,
and Coulomb $z_c = (1431.9 - i 2.4 \times 10^{-4})$ MeV resonances.}
\label{norms2pole.tab}
\begin{tabular}{cccc}
\hline \hline \noalign{\smallskip}
 & $z_1$ &  $z_2$  &  $z_c$  \\
\noalign{\smallskip} \hline \noalign{\smallskip}
$\mathcal{N}_{(\bar{K}N)_{I=0}}$ &
 $1.587 - i 0.4101$  &  $-0.7302 + i 0.5856$  &  $0.500016 - i 0.000018$ \\
$\mathcal{N}_{(\bar{K}N)_{I=1}}$ &
 $0.0003 - i 0.0037$  &  $0.0003 + i 0.0004$  &  $0.499984 + i 0.000017$ \\
$\mathcal{N}_{(\pi \Sigma)_{I=0}}$ &
 $-0.5872 + i 0.4134$  &  $1.730 - i 0.5859$  &  $\sim 10^{-6}$ \\
$\mathcal{N}_{(\pi \Sigma)_{I=1}}$  &
 $0.0002 + i 0.0004$  &  $-0.0001 - i 0.0001$  &  $\sim 10^{-7}$ \\
\noalign{\smallskip} \hline \noalign{\smallskip}
$\mathcal{N}_{0}$  &
 $0.999565 + i 0.003251$  &   $0.999837 - i 0.000330$  &  $0.500016 - i 0.000017$ \\
$\mathcal{N}_{1}$  &
 $0.000435 - i 0.003251$  &  $0.000163 + i 0.000330$  &  $0.499984 + i 0.000017$ \\
\noalign{\smallskip} \hline \noalign{\smallskip}
$\mathcal{N}_{K^- p}$  &
 $0.838 - i 0.253$  &  $-0.367 + i 0.312$  &  $0.999992 + i 3.5 \times 10^{-6}$ \\
$\mathcal{N}_{\bar{K}^0 n}$  &
 $0.749 - i 0.160$  &  $-0.363 + i 0.274$  &  $\sim 10^{-6}$ \\
$\mathcal{N}_{\pi^- \Sigma^+}$  &
 $-0.200 + i 0.152$  &  $0.577 - i 0.209$  &  $\sim 10^{-7}$ \\
$\mathcal{N}_{\pi^0 \Sigma^0}$  &
 $-0.196 + i 0.138$  &  $0.577 - i 0.195$  &  $ \sim 10^{-7}$ \\
$\mathcal{N}_{\pi^+ \Sigma^-}$  &
 $-0.192 + i 0.124$  &  $0.576 - i 0.182$  &  $\sim 10^{-7}$ \\
\noalign{\smallskip} \hline \hline
\end{tabular}
\end{table}
\end{center}
In order to see another effect of isospin non-conservation, we calculated
the norms of our resonant states, which are strictly speaking non-normalizable,
however a regularization procedure and a generalized norm can be defined for
them (see~\cite{norms} and references therein). In our multichannel case the norm
of a resonance wave function $\Psi$ can be written as
\begin{equation}
\label{norm}
\mathcal{N} = ||\Psi|| = \sum_i ||\Psi_{\m{P}_i}|| = \sum_i ||\Psi_{\m{I}_i}||\ ,
\end{equation}
where partial norms $\mathcal{N}_{\m{P}_i}$ ($\mathcal{N}_{\m{I}_i}$) are
\begin{equation}
\label{norm_int}
\mathcal{N}_{\m{P}_i} \equiv ||\Psi_{\m{P}_i}|| = \int\Psi_{\m{P}_i}^2(\vec{k})d\vec{k}.
\end{equation}
Note the square in~(\ref{norm_int}) instead of the modulus squared, due to which the
norms are usually complex. The details of calculating these norms in momentum representation
can be found in~\cite{norms}.
In spite of the fact, that the unique physical interpretation of complex norms
is not completely clear yet, the total wave function $\Psi$ can be normalized
as $\mathcal{N} = ||\Psi|| = 1$ and in this case the partial norms
$\mathcal{N}_{\m{P}_i}$ ($\mathcal{N}_{\m{I}_i}$)
can serve as a measure of contribution of different particle channels to $\Psi$.
The partial norms of our nuclear and Coulomb resonances in both $\m{P}$ and $\m{I}$
representations are shown in Tables~\ref{norms1pole.tab} and~\ref{norms2pole.tab}.

We define the $I=0$ and $I=1$ norms as
\begin{equation}
\mathcal{N}_0 \equiv \mathcal{N}_{(\bar{K}N)_{I=0}} + \mathcal{N}_{(\pi\Sigma)_{I=0}}
\qquad {\rm and} \qquad
\mathcal{N}_1 \equiv \mathcal{N}_{(\bar{K}N)_{I=1}} + \mathcal{N}_{(\pi\Sigma)_{I=1}},
\end{equation}
from the Tables it is seen, that the nuclear resonances are predominantly in
the $I=0$ channel, as expected. The $I=1$ admixture shows up in the fourth digit.
It is also noteworthy, that in the two-pole case
one of the resonances seems to be composed mainly from the $\bar{K}N$ pair,
while the other one from the $\pi\Sigma$. As for the Coulomb level,
again as supposed, it is essentially a $K^-p$ state. The isospin mixing manifests
itself as a small deviation of $\mathcal{N}_{K^-p}$ from unity (or
$\mathcal{N}_{(\bar{K}N)_{I=0}}$ and $\mathcal{N}_{(\bar{K}N)_{I=1}}$ from $0.5$).
We see, that in contrast to the strong scattering length case, isospin breaking
effects play minor role for the resonance wave functions.

\section{Conclusions}

To conclude, we constructed new phenomenological strong isospin-dependent
$\bar{K}N - \pi \Sigma$ potential and investigated the role of isospin breaking effects,
such as direct inclusion of the Coulomb interaction and using physical masses, in the
calculations.
The effects are turned out to be important for the reproducing $1s$ kaonic level shift
and width and for the obtaining the correct $K^- p $  strong scattering length.
We found two ``best'' sets of potential parameters for one-pole and two-pole structure of
$\Lambda(1405)$ resonance describing all experimental data: the level shift and width of
kaonic hydrogen $1s$ level within $1 \sigma$ KEK confidence region, $K^- p$ threshold
branching ratios $\gamma$ and $R_{\pi \Sigma}$, elastic and inelastic $K^- p$ cross-sections,
and $\Lambda(1405)$ resonance shape.
Attempts to move the obtained ($\Gamma$, $\Delta E$) values toward DEAR $1 \sigma$ region
led to drastic worsening of $K^- p$ cross-sections, so we came to the same conclusions,
as~\cite{Borasoy_aKp} that DEAR results are inconsistent with $K^- p$ scattering
data.

Our one- and two-pole ``best'' sets of parameters are of the same quality in describing
existing experimental data. The only large difference between one- and two-pole
variants of the potential is between the kaonic hydrogen widths $\Gamma_{1s}$. However,
even $106$ eV are not sufficient for making conclusions about structure of $\Lambda(1405)$
resonance due to much larger experimental errors of KEK measurement. More precise experimental
data on $K^- p$ atom, for example, from forthcoming SIDDHARTA experiment~\cite{SIDDHARTA}
could choose one of the variants of $\Lambda(1405)$ structure.
More precise data on $K^- p$ cross-sections are also highly desirable.

The corresponding to the potentials $\bar{K}N - \pi \Sigma$ $T$-matrices are suitable and
will be used in a new three-body coupled-channel Faddeev calculation.

\begin{acknowledgments}
The work was supported by the Czech GA AVCR grant KJB100480801.
\end{acknowledgments}

\appendix*
\section{}

The state vector $|\Psi\rangle$ is an element of both configuration and particle space.
In particle space we can use either the particle pair basis $\m{P}$ with elements
$|\m{P}_i\rangle$, $i = 1 \dots 5$:
\begin{equation}
\left[\, |\m{P}_i \rangle \, \right] = \left(
\begin{tabular}{ccccc}
 $|K^-p \rangle$, & $|\bar{K}^0 n \rangle$, & $|\pi^- \Sigma^+ \rangle$, &
$|\pi^0 \Sigma^0 \rangle$, & $|\pi^+ \Sigma^- \rangle$
\end{tabular}
\right)
\end{equation}
or, equivalently, the isospin basis $\m{I}$ with $|\m{I}_i\rangle$, $i = 1 \dots 5$:
\begin{equation}
\left[\, |\m{I}_i \rangle \, \right] = \left(
\begin{tabular}{ccccc}
 $|\bar{K}N \rangle_{I=0}$, & $|\bar{K}N \rangle_{I=1}$, &
 $|\pi \Sigma \rangle_{I=0}$, &
$|\pi \Sigma \rangle_{I=1}$, & $|\pi \Sigma \rangle_{I=2}$
\end{tabular}
\right) \,.
\end{equation}
Here $I$ is a two-particle isospin. The two bases are connected by an orthogonal matrix
composed of the corresponding Clebsch-Gordan coefficients:
\be
|\m{I}_i\rangle = \sum_{j} |\m{P}_j\rangle \langle\m{P}|\m{I}\rangle_{ji}
\ee
with
\be
\label{PIelem}
\langle\m{P}|\m{I}\rangle=\left(\begin{tabular}{ccccc}
$-1/\sqrt{2}$&$1/\sqrt{2}$&$0$&$0$&$0$\\
$1/\sqrt{2}$&$1/\sqrt{2}$&$0$&$0$&$0$\\
$0$&$0$&$1/\sqrt{3}$&$-1/\sqrt{2}$&$1/\sqrt{6}$\\
$0$&$0$&$-1/\sqrt{3}$&$0$&$\sqrt{2/3}$\\
$0$&$0$&$1/\sqrt{3}$&$1/\sqrt{2}$&$1/\sqrt{6}$\\
\end{tabular}\right),  \qquad
\langle\m{I}|\m{P}\rangle = \langle\m{P}|\m{I}\rangle^{\rm T}.
\ee
The projections
\be
\langle\m{P}_i|\Psi\rangle=\Psi_{\m{P}_i}\ \ {\rm{and}}\ \langle\m{I}_i|\Psi\rangle=\Psi_ {\m{I}_i}
\ee
are state vectors in ``ordinary'' space. We can define column vectors
\be
\label{column}
\underline{\Psi}(\m{P})=\{\Psi_{\m{P}_i}\}\ \ {\rm{and}}\ \ \underline{\Psi}(\m{I})=\{\Psi_{\m{I}_i}\}.
\ee
Obviously
\be
\label{PIvec}
\Psi_{\m{I}_i}=\sum_j\langle\m{I}|\m{P}\rangle_{ij}\Psi_{\m{P}_j}\ \ {\rm{or}}\ \
\underline{\Psi}(\m{I})=\langle\m{I}|\m{P}\rangle\ \underline{\Psi}(\m{P}).
\ee
Correspondingly, operators in this case are matrices in particle space with indices according
to the chosen representation:
\be
\underline{\underline{O}}(\m{I}) = [\, O_{\m{I}_i\m{I}_j} \,]\ \ {\rm{or}}\ \
\underline{\underline{O}}(\m{P}) = [\, O_{\m{P}_i\m{P}_j} \,]
\ee
and the matrix elements $O_{\m{I}_i\m{I}_j}$($O_{\m{P}_i\m{P}_j}$) are operators in usual
configuration space.
Again
\be
\label{PImatrix}
O_{\m{I}_i\m{I}_j} =
\sum_{st}\langle\m{I}|\m{P}\rangle_{is}O_{\m{P}_s\m{P}_t}\langle\m{P}|\m{I}\rangle_{tj}
\ \ \ {\rm{or}}\ \ \
\underline{\underline{O}}(\m{I})=\langle\m{I}|\m{P}\rangle\
\underline{\underline{O}}(\m{P})\ \langle\m{P}|\m{I}\rangle .
\ee
Here and in what follows the single and double underlining denotes vectors and matrices in
particle space, respectively.

Our basic operators are $H^0$, $V^c$ and $V^s$, and -- having in mind
Eqs.~(\ref{psiLS})-(\ref{bound1cond}) -- $G^c(z)$.
To define our multichannel problem, we have to specify these operators in particle space.
The operators $H^0$, $V^c$ and $G^c(z)$ do not change the particle composition, therefore
they can be conveniently defined in $\m{P}$ representation, where they are diagonal. Thus
\be
\label{H0}
H^0(\m{P})_{ij} = \delta_{\m{P}_i,\m{P}_j}H^0_{\m{P}_i} =
\delta_{\m{P}_i,\m{P}_j} \left(\frac{\hat{p}^2}{2\mu_{\m{P}_i}}+E_{\m{P}_i}^{th}\right) ,
\ee
where $\mu_{\m{P}_i}$ and $E_{\m{P}_i}^{th}$ are the reduced mass and threshold energy
for the particle pair $\m{P}_i$, respectively, $\hat{p}$ is an operator of relative momentum.
The Coulomb potential acts obviously only
between charged particle pairs, therefore its matrix elements are
\be
V^c(\m{P})_{ij}=\delta_{\m{P}_i,\m{P}_j}V_{\m{P}_i}^c\ ,
\ee
with
\be
V^c_{K^-p} = V^c_{\pi^-\Sigma^+} = V^c_{\pi^+\Sigma^-} = v^c \ \ \rm{and}\ \
V^c_{\bar{K}^0 n} = V^c_{\pi^0\Sigma^0} = 0.
\ee
Here $v^c$ is an ordinary Coulomb potential between two particles with charges $+1$ and $-1$.
Similarly, the corresponding Green's function matrix has the form
\be
\label{Gc}
G^c(\m{P})_{ij}=\delta_{\m{P}_i,\m{P}_j}G^c_{\m{P}_i}(z)\ ,
\ee
with
\be
G^c_{\m{P}_i}(z)=(z-H^0_{\m{P}_i}-V_{\m{P}_i}^c)^{-1}.
\ee
The strong interaction $V^s$, responsible for the transitions between different particle
channels, is supposed to conserve the two-particle isospin $I(\m{I}_i)$, therefore it is
convenient to define it in $\m{I}$ representation. We have chosen a separable form:
\be
V^s(\m{I})_{ij} = \delta_{I(\m{I}_i),I(\m{I}_j)}\
   |g_{\m{I}_i}\rangle\ \lambda_{\m{I}_i,\m{I}_j}\ \langle g_{\m{I}_j}|\ ,
\ee
which can be conveniently rewritten as
\be
V^s(\m{I}) = |\underline{\underline{g}}\rangle \, \underline{\underline{\mathstrut \Lambda}}
\ \langle\underline{\underline{g}}|
\ee
with
\be
| g(\m{I})\rangle_{ij} = \delta_{\m{I}_i,\m{I}_j}\ |g_{\m{I}_i}\rangle
\ee
and
\be
\underline{\underline{\Lambda}}(\m{I})=
\left(\begin{tabular}{ccccc}
$\lambda^0_{\bar{K}N,\bar{K}N}$&$0$&$\lambda^0_{\bar{K}N,\pi\Sigma}$&$0$&$0$\\
$0$&$\lambda^1_{\bar{K}N,\bar{K}N}$&$0$&$\lambda^1_{\bar{K}N,\pi\Sigma}$&$0$\\
$\lambda^0_{\pi\Sigma,\bar{K}N}$&$0$&$\lambda^0_{\pi\Sigma,\pi\Sigma}$&$0$&$0$\\
$0$&$\lambda^1_{\pi\Sigma,\bar{K}N}$&$0$&$\lambda^1_{\pi\Sigma,\pi\Sigma}$&$0$\\
$0$&$0$&$0$&$0$&$\lambda^2_{\pi\Sigma,\pi\Sigma}$\\
\end{tabular}\right)
\ee
(here we moved $I(\m{I}_i) = I(\m{I}_j)$ indices of the matrix elements to the right-up
positions for a convenience).
To complete the description of the matrix-vector analogue of Eqs.(\ref{Tbasc})-(\ref{bound1cond}),
the initial (final) states $|\Phi^c_b\rangle$ have to be specified.
For a given initial (final) particle pair labeled by $\m{P}_i$ the particle space vector
can be conveniently defined in $\m{P}$ representation:
\be
|\Phi^c_{\m{P}_i}(\m{P})_j \rangle =
\delta_{\m{P}_i,\m{P}_j}|\Phi^c_{\m{P}_i}\rangle
\ee
with $|\Phi^c_{\m{P}_i}\rangle$ being an ordinary configuration space state vector
(with the Coulomb interaction taken into account, if it exists for that pair).

Now all operators and states are defined, and we are in position to write down the particle
space matrix analogue of Eq.~(\ref{Tfi}):
\be
\label{Tfi_matrix}
T_{ba}^{sc} = \langle\un{\vphantom{g} \Phi}_{\; b}^{c(-)}|\un{\un{g}} \,
\rangle(\un{\un{\vphantom{g} \Lambda}}^{-1}- \langle \un{\un{g}}
|\un{\un{\vphantom{g} G}}^c(E + i \varepsilon)|\un{\un{g}}\rangle)^{-1} \,
\langle \un{\un{g}}|\un{\vphantom{g} \Phi}_{\; a}^{c(+)}\rangle .
\ee
In the described matrix formulation of the problem the position of the bound states
and resonances instead of Eq.(\ref{bound1cond}) is determined by
\be
\label{boundcplcond}
{\rm Det}(\un{\un{\vphantom{g} \Lambda}}^{-1}-
\langle \un{\un{g}}|\un{\un{\vphantom{g} G}}^c(z)|\un{\un{g}}\rangle) = 0 .
\ee
However, for writing out Eq.~(\ref{Tfi_matrix}) and ~(\ref{boundcplcond}) in components it
is necessary to use the same representation ( $\m{I}$ or $\m{P}$) for all vectors and
matrices. Since we are interested in obtaining parameters of the strong interaction $V_s$
which is given in $\m{I}$, we performed our calculations
in this representation and transformed vectors and matrices defined in $\m{P}$ into
$\m{I}$, using formulae~(\ref{PIelem}), (\ref{PIvec}), and (\ref{PImatrix}). As a non-trivial
example, $\un{\un{G}}^c(\m{I})$ is not a diagonal matrix as $\un{\un{G}}^c(\m{P})$
is~(\ref{Gc}), but has the form:
\be
\label{Gc_matrix}
\un{\un{G}}^c(\m{I})=
\left(\begin{tabular}{cc}
$\un{\un{G^c_{\bar{K}N}}}(\m{I})$&\begin{tabular}{ccc}$0$&$0$&$0$\\$0$&$0$&$0$\end{tabular}\\
\begin{tabular}{cc}$0$&$0$\\$0$&$0$\\$0$&$0$\end{tabular}&$\un{\un{G^c_{\, \pi \Sigma}}}(\m{I})$\\
\end{tabular}\right)
\ee
with
\begin{equation}
\un{\un{G^c_{\bar{K}N}}}(\m{I}) =
\left(\begin{tabular}{cc}
$\frac{1}{2}(G^c_{K^-p}+G^c_{\bar{K}^0n})$&$-\frac{1}{2}(G^c_{K^-p}-G^c_{\bar{K}^0n})$\\
$-\frac{1}{2}(G^c_{K^-p}-G^c_{\bar{K}^0n})$&$\frac{1}{2}(G^c_{K^-p}+G^c_{\bar{K}^0n})$\\
\end{tabular}\right)
\end{equation}
and
\begin{eqnarray}
&& \hspace{-4mm}
\un{\un{G^c_{\, \pi \Sigma}}}(\m{I}) =  \\
\nonumber
&&  \hspace{-4mm}
\left(\begin{tabular}{ccc}
$\frac{1}{3}(G^c_{\pi^-\Sigma^+}+G^c_{\pi^0\Sigma^0}+G^c_{\pi^+\Sigma^-})$&
$-\frac{1}{\sqrt{6}}(G^c_{\pi^-\Sigma^+}-G^c_{\pi^+\Sigma^-})$&
$\frac{1}{3\sqrt{2}}(G^c_{\pi^-\Sigma^+}-2G^c_{\pi^0\Sigma^0}+G^c_{\pi^+\Sigma^-})$\\
$-\frac{1}{\sqrt{6}}(G^c_{\pi^-\Sigma^+}-G^c_{\pi^+\Sigma^-})$&
$\frac{1}{2}(G^c_{\pi^-\Sigma^+}+G^c_{\pi^+\Sigma^-})$&
$-\frac{1}{2\sqrt{3}}(G^c_{\pi^-\Sigma^+}-G^c_{\pi^+\Sigma^-})$\\
$\frac{1}{3\sqrt{2}}(G^c_{\pi^-\Sigma^+}-2G^c_{\pi^0\Sigma^0}+G^c_{\pi^+\Sigma^-})$&
$-\frac{1}{2\sqrt{3}}(G^c_{\pi^-\Sigma^+}-G^c_{\pi^+\Sigma^-})$&
$\frac{1}{6}(G^c_{\pi^-\Sigma^+}+4G^c_{\pi^0\Sigma^0}+G^c_{\pi^+\Sigma^-})$
\end{tabular}\right)
\end{eqnarray}
It can be seen, that  $\un{\un{G}}^c(\m{I})$ has matrix elements connecting states with
unequal isospins. They are proportional to difference of $\un{\un{G}}^c$ components
in dissimilar particle pair channels $\m{P}_i$. This isospin non-conservation has two
independent sources. First, $G^c$ of charged particles differs from $G^c = G^0$ of neutral pairs,
second, due to the mass difference of the isomultiplet members, the particle pairs
$\m{P}_i$ have different reduced masses and threshold energies, and thus, according to
Eq.~(\ref{H0}), different $H^0$-s and $G^0$-s. Neglecting these two effects in the
$\pi\Sigma$ sector leads to a diagonal submatrix $\un{\un{G^c_{\, \pi \Sigma}}}(\m{I})$.

\end{document}